\documentclass{moriond}

\usepackage{amsmath}
\usepackage{amssymb}
\usepackage{multirow}

\def\be{\begin{equation}}
\def\ee{\end{equation}}
\def\bea{\begin{eqnarray}}
\def\eea{\end{eqnarray}}

\newcommand{\ie}{{\it i.e.}}

\newcommand{\Sbar}{S^\dagger}
\newcommand{\Qmax}{Q_{\rm max}}
\newcommand{\Qmin}{Q_{\rm min}}

\newcommand{\lc}{\lambda_{\phi S}}

\newcommand{\lphi}{\lambda_\phi}

\newcommand{\ls}{\lambda_S}
\newcommand{\ms}{m_{S,0}}

\newcommand{\Photo}{}

\begin{document}
\vspace*{4cm}
\title{Origin and detection of nontopological soliton dark matter}

\author{Nicholas Orlofsky}

\address{Institute of Theoretical Physics, Faculty of Physics, University of Warsaw, ul. Pasteura 5, \\ PL-02-093 Warsaw, Poland}

\maketitle\abstracts{
Macroscopic dark matter like nontopological solitons can form either via the fusion and accumulation of free particles or during cosmological phase transitions. Both mechanisms can create dark matter with large masses ranging from TeV to solar mass. This can lead to interesting targets in direct detection, astrophysical, and cosmological searches.
}

\section{Introduction to scalar nontopological solitons}

Simple models of dark matter (DM) can contain attractive interactions that lead to the formation of macroscopic compact objects. The focus of this contribution is how such structures form, how large they can be, whether they can dominate the dark sector, and how they can be detected. 
Many of the results of this contribution can be found in Ref.~\cite{Bai:2022kxq}, where more details are provided.

Scalar nontopological solitons (NTSs) or Q-balls \cite{Coleman:1985ki} provide a representative for how various types of DM models form compact objects. Additionally, they are extremely simple, requiring as few as one new field beyond the Standard Model (SM) \cite{Ponton:2019hux}. Perhaps the simplest renormalizable theory involves a complex scalar $S$ with a $U(1)_S$ global symmetry and another scalar $\phi$ with any symmetry that only allows it to enter the potential in even powers. Then, the scalar potential is $V(S, \phi) = \frac{1}{4} \lphi (|\phi|^2 - v^2)^2 +  \frac{1}{4} \lc |S|^2 |\phi|^2 + \ls |S|^4 + \ms^2 |S|^2$. Nontrivial variations on this include having $\phi$ be the SM Higgs \cite{Ponton:2019hux}, having $\phi$ symmetry breaking lead to topological charges \cite{Bai:2021mzu}, or having the $S$ symmetry be gauged \cite{Lee:1988ag}. The NTS solution is obtained with the radially symmetric ansatz $S=e^{-i \omega t}\,v\,s(r)/\sqrt{2}$ and $\phi = v\, f(r)$, with $\omega$ a free parameter that turns out to set the NTS charge and $r$ and $t$ the radial and time coordinates. The classical equations of motion for $s(r)$ and $f(r)$ lead to solutions where the $\phi$ symmetry is partially or fully restored and the $S$ charge is nonzero near the origin. These solutions can be thought of as trading off an increase in the $\phi$ vacuum energy with a decrease in the $S$ mass. At large charge, the NTS mass and radius are related to its charge $Q$ by $m_{Q} \approx  v Q [\left(\ls \lphi\right)^{1/4} + c_2 Q^{-1/3} ]$ and 
$R_{Q} \approx \frac{3^{1/3} \ls^{1/12} }{(4\pi)^{1/3}\lphi^{1/4} v} Q^{1/3}$, 
respectively, with $c_2$ a numerical factor. Thus, large-charge NTSs are stable provided $m_S^2 = \frac{1}{4} \lc v^2 + \ms^2 < v^2 \left(\ls \lphi\right)^{1/2}$. At small charge, the mass-to-charge ratio increases with decreasing $Q$ until $m_Q/Q > m_S$, leading to a minimum stable charge $\Qmin$.

For the sake of comparison, 
we contrast the prior model, which we dub ``Model B,'' with 
the one used in Refs.~\cite{Frieman:1988ut,Griest:1989bq}, or ``Model A.'' In Model A, $V(S, \sigma) = \frac{1}{8} \lambda\, (\sigma^2 - \sigma_0^2)^2 + \frac{1}{3} \lambda_2\, \sigma_0 (\sigma-\sigma_0)^3 +  \frac{m_S^2}{(\sigma_- - \sigma_0)^2} |S|^2 (\sigma - \sigma_0)^2 + \Lambda$, with $\sigma$ a real scalar field. This is essentially the same model, but with no quartic $\lambda_S$ term. Without this self-repulsive coupling, the NTS is more compact than in Model B with properties $m_Q = 5.15 \sigma_0 \lambda^{1/4} Q^{3/4}$ and $R_Q = 0.8 \lambda^{-1/4} \sigma_0^{-1} Q^{1/4}$. 

\section{Solitosynthesis}

Solitosynthesis is the process of NTS formation via the fusion and accumulation of free particles. NTSs of charge $Q$ denoted by $(Q)$ can interact with free particles in interactions including $(Q) + S \leftrightarrow (Q+1) + X,$ $(Q) + \Sbar \leftrightarrow (Q-1) + X,$ and $(\Qmin) + \Sbar \leftrightarrow  S + S + \cdots +S + X$ with $S$ appearing $\Qmin-1$ times on the right hand side. $X$ denotes other degrees of freedom to conserve energy and momentum. For simplicity, we assume the radiative capture of a free particle on a NTS of charge $Q$ has a cross section $(\sigma v_{\rm rel})_Q = \pi R_Q^2$ \cite{Bai:2019ogh}, and the rate of the inverse process is obtained by detailed balance. Interaction rates between NTSs are negligible in the scenarios considered here. Free $S$ and $S^\dagger$ particles can also annihilate, including to $\phi + \phi^\dagger$. All of these interactions lead to equilibrium number densities $n_{i=S,\Sbar,Q,-Q}^{\rm eq}$ with a common chemical potential $\mu = \mu_S = -\mu_{\Sbar} = \mu_Q / Q$. For the NTS abundance not to be suppressed, an initial $S$ particle-antiparticle asymmetry $\eta \equiv (n_S - n_\Sbar)/n_\gamma$ must be present, where $n_\gamma$ is the SM photon number density. This uniquely sets the value of $\mu$, and thus 
$n_i^{\rm eq}$, as a function of temperature. 

In equilibrium, at high temperatures free $S$ particles dominate because it is very easy to knock a particle loose from a NTS. At intermediate temperatures around the free particle mass, the antiparticles and anti-NTSs will be annihilated away. Finally, at temperatures well below the free particle mass, almost all of the $S$ charge will reside in NTSs with the largest attainable charge $\Qmax$ because this is the lowest possible energy state. $\Qmax$ can be estimated by demanding that the time for a NTS to accumulate charge from $\Qmin$ to $\Qmax$ be less than about a Hubble time: $\tau_{\Qmin \to \Qmax} = \sum_{Q=\Qmin}^{\Qmax} (n_S \, (\sigma v_{\rm rel})_Q)^{-1} \lesssim H^{-1}$, with $H$ the Hubble parameter.

To make further progress, analytic estimates must be employed because a full numeric result would require solving $\mathcal{O}(\Qmax)$ coupled Boltzmann equations. The charge-domination temperature $T_D$ can be analytically estimated from $n_S^\text{eq} \approx \Qmax n_{\Qmax}^\text{eq} \approx \eta n_\gamma / 2$.
Because $m_S < m_Q/Q$ for stable NTSs, 
$T_D$ is larger than the energy-domination temperature $T_\rho$
when $m_S n_{S} \approx m_{\Qmax} n_{\Qmax}$.

A key finding of our work is an improved estimate of the freeze-out temperature $T_F$ compared to prior works \cite{Griest:1989bq,Frieman:1989bx,Postma:2001ea}. If the Boltzmann equations for all NTSs are summed (ignoring anti-NTSs), then the Boltzmann equation for the summed NTS number density $n_\text{NTS} \equiv \sum_{Q=\Qmin}^{\Qmax} n_{Q}$ is $\dot{n}_\text{NTS} + 3\,H\,n_\text{NTS} = - (\sigma v_{\rm rel})_{Q_\text{min}} [n_{\Qmin} n_{\Sbar} - n_{\Qmin}^\text{eq} n_{\Sbar}^\text{eq} ( n_S/n_S^\text{eq} )^{Q_\text{min} - 1} ]$. Then, the freeze-out temperature can be estimated by $H\,n_\text{NTS}^\text{eq} \sim (\sigma v_{\rm rel})_{Q_\text{min}} n_{\Qmin}^\text{eq} n_{\Sbar}^\text{eq} \left| \right. _{T=T_F}$. If $T_F<T_D$, $n_\text{NTS}^\text{eq} \approx n_{\Qmax}^\text{eq}$ on the left hand side, which leads to a simple analytic expression for $T_F$. For the case $T_F > T_D$, $n_{\Qmax}^\text{eq} \leq n_\text{NTS}^\text{eq}$, so the analytic estimate will not give the correct value of $T_F$. However, it will correctly indicate that $T_F >T_D$, \ie, that free particles dominate NTSs.

\begin{figure}[t!]
	\centering
	\includegraphics[width=0.38\textwidth]{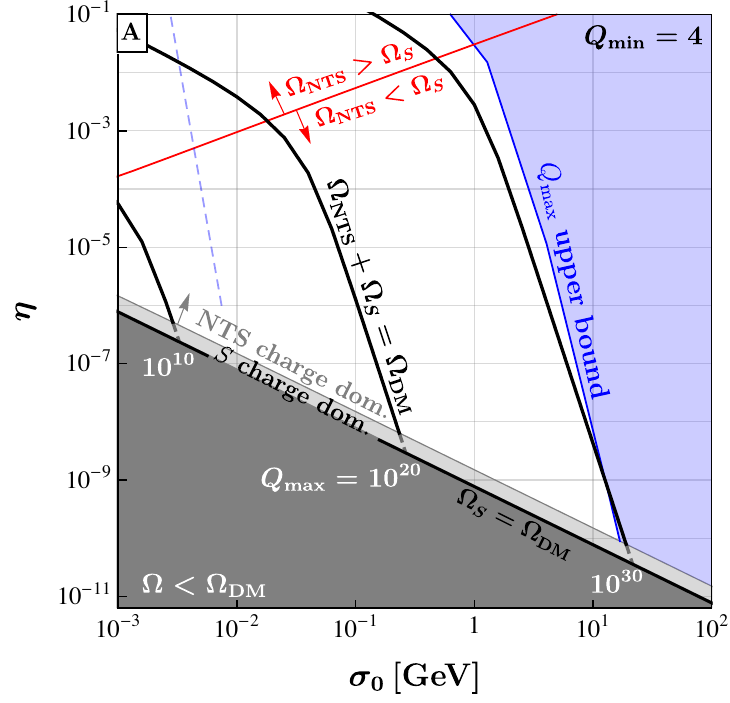}
	\includegraphics[width=0.37\textwidth]{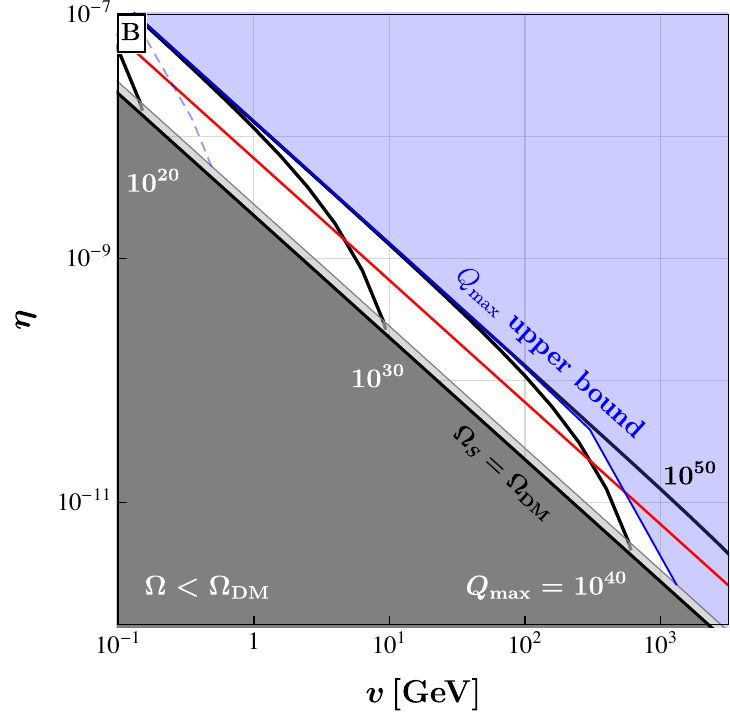}
	\caption[]{
		Parameter space where NTSs and free $S$ particles make up all of DM after solitosynthesis for Models A (left) and B (right).
		The gray and red lines separate the NTS and $S$-particle charge- and energy-dominated regions, respectively.
		Black solid curves correspond to different choices of $\Qmax$. 
		Right of the blue solid (dashed) line is excluded by the upper bound on $\Qmax$ for $\Qmin=4~(7)$.
		Model parameters are taken as in Table~\ref{tab:benchmarks} (except Model B $\Qmin=7$ for which $\lc=6.2$). Reproduced from Ref.~\cite{Bai:2022kxq}. 
	}
	\label{fig:relic_abundance_SS}
\end{figure}

With these analytic estimates for $T_D$, $T_\rho$, $T_F$, and $\Qmax$, one can determine the necessary values of $\eta$ and the NTS model properties (such as $\Qmin$ and 
the symmetry breaking scale $\sigma_0$ or $v$) for NTSs to dominate, as well as the typical relic NTS properties. This parameter space is shown in Fig.~\ref{fig:relic_abundance_SS}, and examples assuming $\Qmax$ saturates its upper bound are shown in Table~\ref{tab:benchmarks}.

\section{Phase transition formation}

NTSs could also form during the cosmological phase transition (PT) of the $\phi$ or $\sigma$ field. 
The $S$ particles energetically prefer to remain in false vacuum regions where the $\phi$ or $\sigma$ symmetry is restored and the $S$ mass is lower, which can lead to the formation of NTSs. A feature of this formation mechanism is that NTSs can form even with $\eta=0$ due to statistical fluctuations in the numbers of $S$ and $\Sbar$ particles within a given false vacuum region. Specifically, if false vacuum regions
during a PT
contain a typical total $N_S$ of $S$ and $\Sbar$ particles, then the typical NTS charge is $\langle Q \rangle \sim \max [\eta N_S, \, N_S^{1/2} ]$. Details of the calculation are in Ref.~\cite{Bai:2022kxq}. If $\phi$ spontaneous symmetry breaking produces topological monopoles, they will tend to form inside of NTSs \cite{Bai:2021mzu}.

The NTSs formed during a PT may be modified by later solitosynthesis processes. If 
the PT temperature
$T_f>T_F$ and 
chemical equilibrium is reached, then the initial NTSs have no effect on the final abundances and charge distributions. On the other hand, if 
$m_S/T_f \gtrsim 50$, then free $S$ particles are 
not thermally produced, able to escape false vacuum regions, or dislodged from NTSs
in sufficient quantities to modify the NTS distribution formed during the PT.

\begin{table}[t]
\small
	\centering
	\begin{tabular}{l| l|l||l | l| l |l}
		\hline \hline
		Mechanism & Model & $\eta$ & $m_Q$ (g) & $R_Q$ (m) & $\langle Q \rangle$ & $\sigma_0$  or $v$ 
		(GeV) \\ \hline
		\multirow{3}{*}{Solitosynthesis} & A & $10^{-10}$ & 3 & $3 \times 10^{-10}$ & $6 \times 10^{29}$ & 10 \\
		& B & $10^{-10}$ & $5\times 10^{22}$ & $2\times 10^{-3}$ & $1 \times 10^{45}$ & $1\times 10^2$ \\
		& B & $10^{-6}$ & $6\times 10^{30}$ & $3\times 10^5$ & $1\times 10^{57}$ & $1 \times 10^{-2}$ \\\hline
		\multirow{3}{*}{First order PT} & A & 0 & $9\times 10^{-6}$ & $5\times 10^{-23}$ & $3 \times 10^{11}$ & $2\times 10^9$ \\
		& B & 0 & $2\times 10^{-3}$ & $4 \times 10^{-19}$ & $1 \times 10^{14}$ & $4 \times 10^7$ \\ 
		& B & $10^{-4}$ & $8 \times 10^{26}$ & $1 \times 10^{5}$ & $7 \times 10^{53}$ & $3 \times 10^{-3}$ \\\hline
		\multirow{2}{*}{Second order PT} & A & 0 & $2 \times 10^{-20}$ & $3\times 10^{-15}$ & $5 \times 10^4$ & $7 \times 10^{-1}$ \\
		& B & 0 & $1 \times 10^{-20}$ & $5 \times 10^{-13}$ & $5 \times 10^4$ & $2 \times 10^{-2}$ \\
		\hline \hline
	\end{tabular}
	\caption[]{
		Benchmark relic NTS properties in Models A and B as produced from solitosynthesis  or a PT making up all of dark matter (except solitosynthesis of Model A, see text). Model parameters are chosen as $\lambda=1 ,  \, \lambda_2=0.15 \lambda, \,\Qmin=4$ in A and $\lambda_\phi=0.01\, , \lambda_{\phi S}=10, \, \lambda_S=0.2, \, \ms=0$ in B (which gives $\Qmin=4$), except the second order PT where $\lambda=\lphi=10^{-3}$ is used.  Reproduced from Ref.~\cite{Bai:2022kxq}. 
	} 
	\label{tab:benchmarks}
\end{table}

Table \ref{tab:benchmarks} shows examples of NTS properties from solitosynthesis and PTs. In all examples, 100\% of DM is in NTSs and free $S$ particles, and NTSs dominate the DM energy density, except for solitosynthesis of Model A where NTSs only dominate the $S$ charge density. Notice that $\eta=10^{-10}$ (motivated by the SM asymmetry) or $\eta=0$ can produce macroscopically large NTSs, and bigger $\eta$ leads to larger NTSs. Solitosynthesis tends to produce larger NTSs than PTs.

\section{Detection strategies}

There are many possible gravitational probes of macroscopic DM. Existing microlensing and accretion searches constrain DM masses $\gtrsim 10^{23}~\text{g}$ from making up more than $\mathcal{O}(1~\text{to}~10\%)$ of DM, depending on its mass distribution \cite{Carr:2020xqk,Bai:2020jfm}. Very high energy compact sources such as X-ray pulsars \cite{Bai:2018bej} and gamma ray bursts \cite{Katz:2018zrn} may be used 
for 
future lensing searches at lower masses. Gravitational wave searches can also probe lower masses if the dark sector has additional forces~\cite{Bai:2023lyf}. DM masses near the Planck scale could 
even be probed directly by mechanical sensor arrays~\cite{Windchime:2022whs}. 

If DM has additional nongravitational interactions with the SM, masses between $10^8$ and $10^{24}~\text{GeV}$ can be probed by direct detection or neutrino experiments searching for multiple detector hits  \cite{Ponton:2019hux,Bai:2019ogh,DEAPCollaboration:2021raj,Bai:2022nsv}. Planetary-radius DM could be detected as transiting exoplanets \cite{Bai:2023mfi}. 

To conclude, NTSs are a relatively generic prediction when new scalar fields are added to the SM. Further, they would be expected to form in the early universe, often with macroscopic sizes and making up all of DM. Because macroscopic DM can form so readily in simple extensions of the SM such as this, it is worthwhile to continue detection efforts.

\section*{Acknowledgments}

The author thanks Yang Bai and Sida Lu for collaboration on the results presented in this contribution and the Moriond organizers and staff. The work of NO is supported by the National Science Centre, Poland, under research grant no.~2020/38/E/ST2/00243.

\section*{References}

\bibliography{NTSorigin}

\begin{thebibliography}{10}

\bibitem{Bai:2022kxq}
Yang Bai, Sida Lu, and Nicholas Orlofsky.
\newblock {Origin of nontopological soliton dark matter: solitosynthesis or phase transition}.
\newblock {\em JHEP}, 10:181, 2022.

\bibitem{Coleman:1985ki}
Sidney~R. Coleman.
\newblock {Q-balls}.
\newblock {\em Nucl. Phys. B}, 262(2):263, 1985.
\newblock [Addendum: Nucl.Phys.B 269, 744 (1986)].

\bibitem{Ponton:2019hux}
Eduardo Pont\'on, Yang Bai, and Bithika Jain.
\newblock {Electroweak Symmetric Dark Matter Balls}.
\newblock {\em JHEP}, 09:011, 2019.

\bibitem{Bai:2021mzu}
Yang Bai, Sida Lu, and Nicholas Orlofsky.
\newblock {Q-monopole-ball: a topological and nontopological soliton}.
\newblock {\em JHEP}, 01:109, 2022.

\bibitem{Lee:1988ag}
Ki-Myeong Lee, Jaime~A. Stein-Schabes, Richard Watkins, and Lawrence~M. Widrow.
\newblock {Gauged q Balls}.
\newblock {\em Phys. Rev. D}, 39:1665, 1989.

\bibitem{Frieman:1988ut}
Joshua~A. Frieman, G.~B. Gelmini, Marcelo Gleiser, and Edward~W. Kolb.
\newblock {Solitogenesis: Primordial Origin of Nontopological Solitons}.
\newblock {\em Phys. Rev. Lett.}, 60:2101, 1988.

\bibitem{Griest:1989bq}
Kim Griest and Edward~W. Kolb.
\newblock {Solitosynthesis: Cosmological Evolution of Nontopological Solitons}.
\newblock {\em Phys. Rev. D}, 40:3231, 1989.

\bibitem{Bai:2019ogh}
Yang Bai and Joshua Berger.
\newblock {Nucleus Capture by Macroscopic Dark Matter}.
\newblock {\em JHEP}, 05:160, 2020.

\bibitem{Frieman:1989bx}
Joshua~A. Frieman, Angela~V. Olinto, Marcelo Gleiser, and Charles Alcock.
\newblock {Cosmic Evolution of Nontopological Solitons. 1.}
\newblock {\em Phys. Rev. D}, 40:3241, 1989.

\bibitem{Postma:2001ea}
Marieke Postma.
\newblock {Solitosynthesis of Q balls}.
\newblock {\em Phys. Rev. D}, 65:085035, 2002.

\bibitem{Carr:2020xqk}
Bernard Carr and Florian Kuhnel.
\newblock {Primordial Black Holes as Dark Matter: Recent Developments}.
\newblock {\em Ann. Rev. Nucl. Part. Sci.}, 70:355--394, 2020.

\bibitem{Bai:2020jfm}
Yang Bai, Andrew~J. Long, and Sida Lu.
\newblock {Tests of Dark MACHOs: Lensing, Accretion, and Glow}.
\newblock {\em JCAP}, 09:044, 2020.

\bibitem{Bai:2018bej}
Yang Bai and Nicholas Orlofsky.
\newblock {Microlensing of X-ray Pulsars: a Method to Detect Primordial Black
  Hole Dark Matter}.
\newblock {\em Phys. Rev. D}, 99(12):123019, 2019.

\bibitem{Katz:2018zrn}
Andrey Katz, Joachim Kopp, Sergey Sibiryakov, and Wei Xue.
\newblock {Femtolensing by Dark Matter Revisited}.
\newblock {\em JCAP}, 12:005, 2018.

\bibitem{Bai:2023lyf}
Yang Bai, Sida Lu, and Nicholas Orlofsky.
\newblock {Gravitational Waves From More Attractive Dark Binaries}.
\newblock [arXiv:2312.13378].

\bibitem{Windchime:2022whs}
Alaina Attanasio et~al.
\newblock {Snowmass 2021 White Paper: The Windchime Project}.
\newblock In {\em {2022 Snowmass Summer Study}}, 3 2022.

\bibitem{DEAPCollaboration:2021raj}
P.~Adhikari et~al.
\newblock {First Direct Detection Constraints on Planck-Scale Mass Dark Matter
  with Multiple-Scatter Signatures Using the DEAP-3600 Detector}.
\newblock {\em Phys. Rev. Lett.}, 128(1):011801, 2022.

\bibitem{Bai:2022nsv}
Yang Bai, Joshua Berger, and Mrunal Korwar.
\newblock {IceCube at the frontier of macroscopic dark matter direct
  detection}.
\newblock {\em JHEP}, 11:079, 2022.

\bibitem{Bai:2023mfi}
Yang Bai, Sida Lu, and Nicholas Orlofsky.
\newblock {Dark exoplanets}.
\newblock {\em Phys. Rev. D}, 108(10):103026, 2023.

\end{thebibliography}

\end{document}